\documentclass{jps-cp}

\usepackage{graphicx}
\usepackage{color}
\usepackage{wrapfig}
\title{Field-Induced Spin Nematic Liquid of the $S=1/2$ Bond-Alternating Chain with the Anisotropy}

\author{Ryosuke~\textsc{Nakanishi}$^{1}$, Takaharu~\textsc{Yamada}$^{1}$, 
Rito~\textsc{Furuchi}$^1$, Hiroki~\textsc{Nakano}$^1$, Hirono~\textsc{Kaneyasu}$^1$,
Kiyomi~\textsc{Okamoto}$^1$, Takashi~\textsc{Tonegawa}$^{1,2,3}$, and T\^oru~\textsc{Sakai}$^{1,4}$}

\inst{
$^{1}$Graduate School of Science, University of Hyogo, Hyogo 678-1297, Japan \\
$^{2}$Professor Emeritus, Kobe University, Kobe 657-8501, Japan\\
$^{3}$Department of Physics, Graduate School of Science, Osaka Metropolitan University, Sakai, Osaka 599-8531, Japan\\
$^{4}$National Institutes for Quantum Science and Technology (QST), SPring-8, Hyogo 679-5148, Japan
}

\email{ryosukenakanishi7@gmail.com}

\recdate{July 12, 2022}

\abst{
The $S=1/2$ ferromagnetic-antiferromagnetic bond-alternating spin chain with the anisotropy
on the ferromagnetic exchange interaction in 
magnetic field is investigated using the numerical diagonalization and the density matrix renormalization 
group analyses. 
It is found that the nematic-spin-dominant Tomonaga-Luttinger liquid phase is induced by the 
external magnetic field for sufficiently large anisotropy. 
The phase diagram with respect to the anisotropy and the magnetization is presented. 
}

\kword{quantum spin system, spin nematic liquid, quantum spin liquid, quantum phase transition}

\def\Vec#1{\mbox{\boldmath $#1$}}

\begin{document}
\maketitle

\section{Introduction}

The spin nematic order \cite{nematic1,nematic2} is one of interesting topics in the field of the low-temperature physics. 
It is the long-range quadrupole order of spins by forming the two-magnon bound state. 
It is a kind of an intermediate state between the conventional long-range magnetic order 
and the quantum spin liquid. 
The previous theoretical mechanisms of the spin nematic order have been 
based on the biquadratic exchange 
interaction\cite{chubukov, solyom, lauchli, senthil, manmana, mao, fridman} 
which directly stabilizes the 
nematic correlation, or the spin 
frustration\cite{chandra,nakatsuji,tsunetsugu,penc,shenoy,park,chubukov2,vekua,sudan,hikihara} 
which suppresses the conventional 
long-range order. 
We note that we showed the existence of the spin nematic liquid phase
under zero  magnetic field in a model having neither the biquadratic interaction nor
the frustration \cite{tone2018}.
Also we found the spin nematic liquid phase under magnetic field in several one-dimensional models
\cite{sakai2010,sakai2020,sakai2021,sakai2022,sakai2022a}.
In this paper we propose a simple model that possibly exhibits the spin nematic liquid phase 
in magnetic field without the biquadratic interaction or the frustration. 
Our present model is the $S=1/2$ ferromagnetic-antiferromagnetic bond-alternating spin chain 
with the Ising-like coupling anisotropy at the ferromagnetic bond. 
When the external magnetic field is applied to this system, 
the gapless Tomonaga-Luttinger liquid (TLL) phase is expected to be realized. 
The present numerical diagonalization analysis indicates that 
the conventional TLL (CTLL) phase would changes to another TLL phase where 
the quasiparticle is the two-magnon bound state for sufficiently 
large anisotropy. 
The present analysis of the critical exponents of some 
spin correlation functions reveals that the nematic-spin-correlation 
dominant TLL region appears in the two-magnon TLL phase. 
A typical phase diagram with respect to the anisotropy and the 
magnetization is presented.

\section{Model}

We consider the magnetization process 
of the $S=1/2$ ferromagnetic-antiferromagnetic bond-alternating spin chain 
with the Ising-like coupling anisotropy at the ferromagnetic bond. 
The Hamiltonian is given by 
\begin{eqnarray}
\label{ham}
&{\cal H}&={\cal H}_0+{\cal H}_Z, \\
&{\cal H}_0& = J_1 \sum_{j=1}^L\left[S_{2j-1}^xS_{2j}^x + S_{2j-1}^yS_{2j}^y  + \lambda S_{2j-1}^zS_{2j}^z  \right] 
 +J_2  \sum_{j=1}^L \Vec{S}_{2j}\cdot \Vec{S}_{2j+1}  \\
&{\cal H}_{\rm Z}& =-H \sum_{j=1}^L \left( S_{2j-1}^z +S_{2j}^z \right),
\end{eqnarray}
where $\lambda$ is a coupling anisotropy parameter  
and $H$ is the external magnetic field along the $z$ direction. 
The ferromagnetic interaction constant $J_1$ is set to $-1$. 
We consider the case in which $J_2>0$ (antiferromagnetic) and 
the anisotropy of the $J_1$ bond is Ising-like ($\lambda >1$). 
For the length $L$ system, 
the lowest energy of ${\cal H}_0$ in the subspace
$\sum _j S_j^z=M$ is denoted by $E(L,M)$. 
The reduced magnetization $m$ is defined by $m=M/M_{\rm s}$, 
where $M_{\rm s}$ denotes the saturation of the magnetization, 
namely $M_{\rm s}=L$. 
The energy $E(L,M)$ is calculated by the Lanczos algorithm under the 
periodic boundary condition ($ \Vec{S}_{2L+1}=\Vec{S}_{1}$). 
Our calculation of the magnetization curve indicates that 
for $J_2 \gtrsim |J_1|$ the spin nematic liquid phase 
does not appear, because the magnetization jump like the 
spin flop occurs. 
Thus in this paper we fix $J_2$ to 0.3, which is a typical case where the 
spin nematic liquid appears clearly.

\section{Ground State without Magnetic Field}

\begin{figure}[h]
   \begin{minipage}{.48\linewidth}
      \begin{center}
         \includegraphics[scale=0.25]{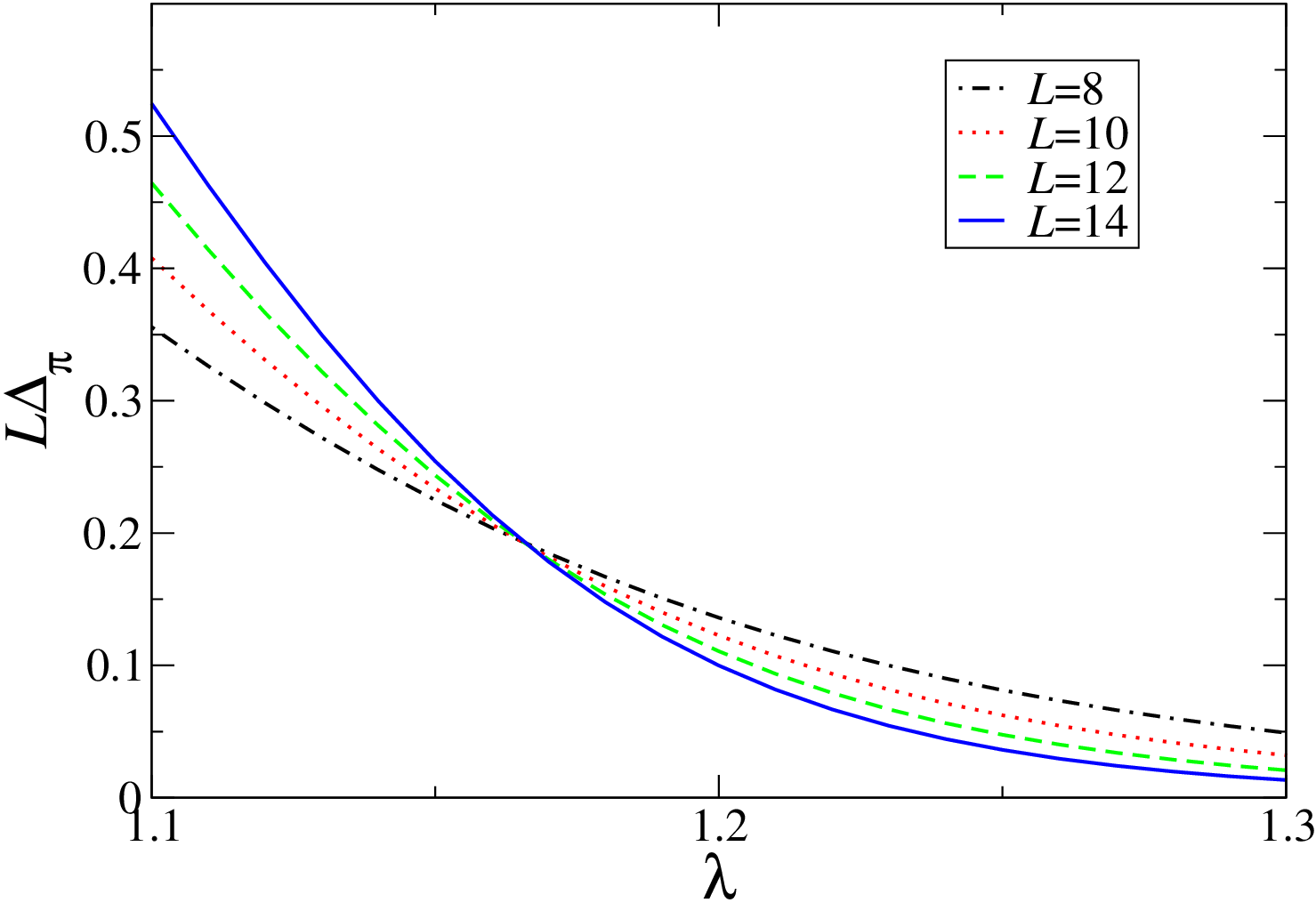}
         \caption{Scaled gap $L\Delta_{\pi}$ is plotted versus $\lambda$ for $L=8, 10, 12, 14$ when $m=0$.
         \par
         ~\par
         ~\par
         ~\par}
         \label{prg}
      \end{center}
   \end{minipage}
   \hspace{0.04\columnwidth}
   \begin{minipage}{.48\linewidth}
      \begin{center}
        \includegraphics[scale=0.25]{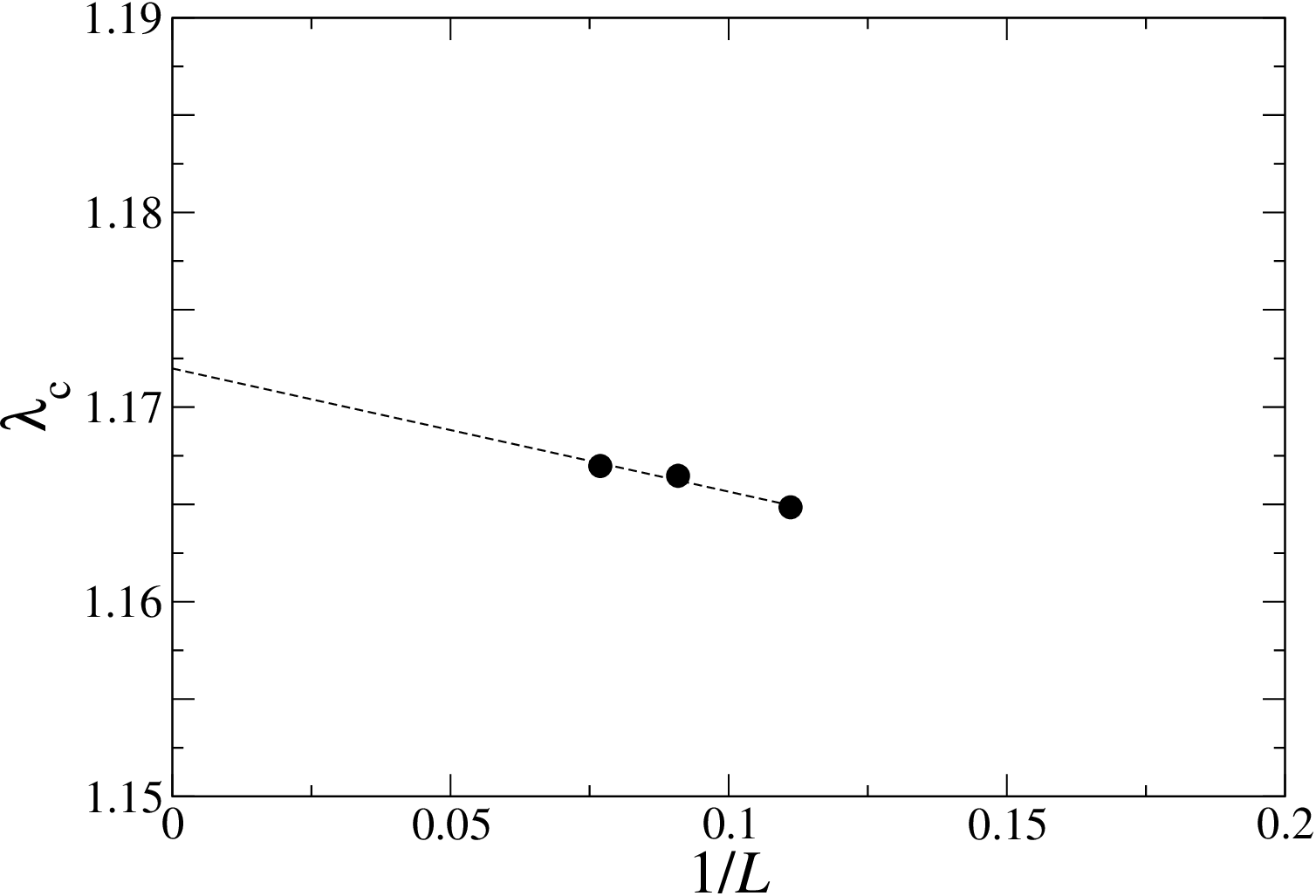}
        \caption{Extrapolation of the size-dependent fixed point $\lambda_{\rm c}(L+1)$ for $L$ and $L+2$ 
           assuming the size correction proportional to $1/(L+1)$. 
           It results in $\lambda _{\rm c}=1.172 \pm 0.001$ in the infinite length limit. 
           }
        \label{extrap}
      \end{center}
   \end{minipage}
\end{figure}

The ground state of the ferromagnetic and antiferromagnetic bond-alternating chain without 
magnetic field is in the Haldane phase when $\lambda =1$,
whereas it would be in the N\'eel ordered phase for 
sufficiently large $\lambda$. 
The phase boundary between these two phases can be estimated with the phenomenological 
renormalization \cite{nightingale}. 
The size-dependent phase boundary is estimated by the fixed point equation for the two system 
sizes $L$ and $L+2$ 
\begin{eqnarray}
L\Delta_{\pi}(L,\lambda)=(L+2)\Delta _{\pi}(L+2,\lambda),
\label{eq:prg}
\end{eqnarray}
where $\Delta_{\pi}$ is the excitation gap with $k=\pi$ in 
the subspace with $M=0$. 
The scaled gap $L\Delta_{\pi}$ is plotted versus $\lambda$ for $L=8, 10, 12, 14$ 
in Fig. \ref{prg}. 
It suggests that the phase boundary exists around $\lambda \sim 1.17$. 
The extrapolation of the size-dependent fixed point $\lambda_{\rm c}(L+1)$ for $L$ and $L+2$ 
assuming the size correction proportional to $1/(L+1)$, as shown in Fig. \ref{extrap}, 
results in $\lambda _{\rm c}=1.172 \pm 0.001$ in the infinite length limit. 

\section{Two Tomonaga-Luttinger Liquids}

\begin{figure}[h]
   \begin{minipage}{.48\linewidth}
      \begin{center}
         \includegraphics[scale=0.25]{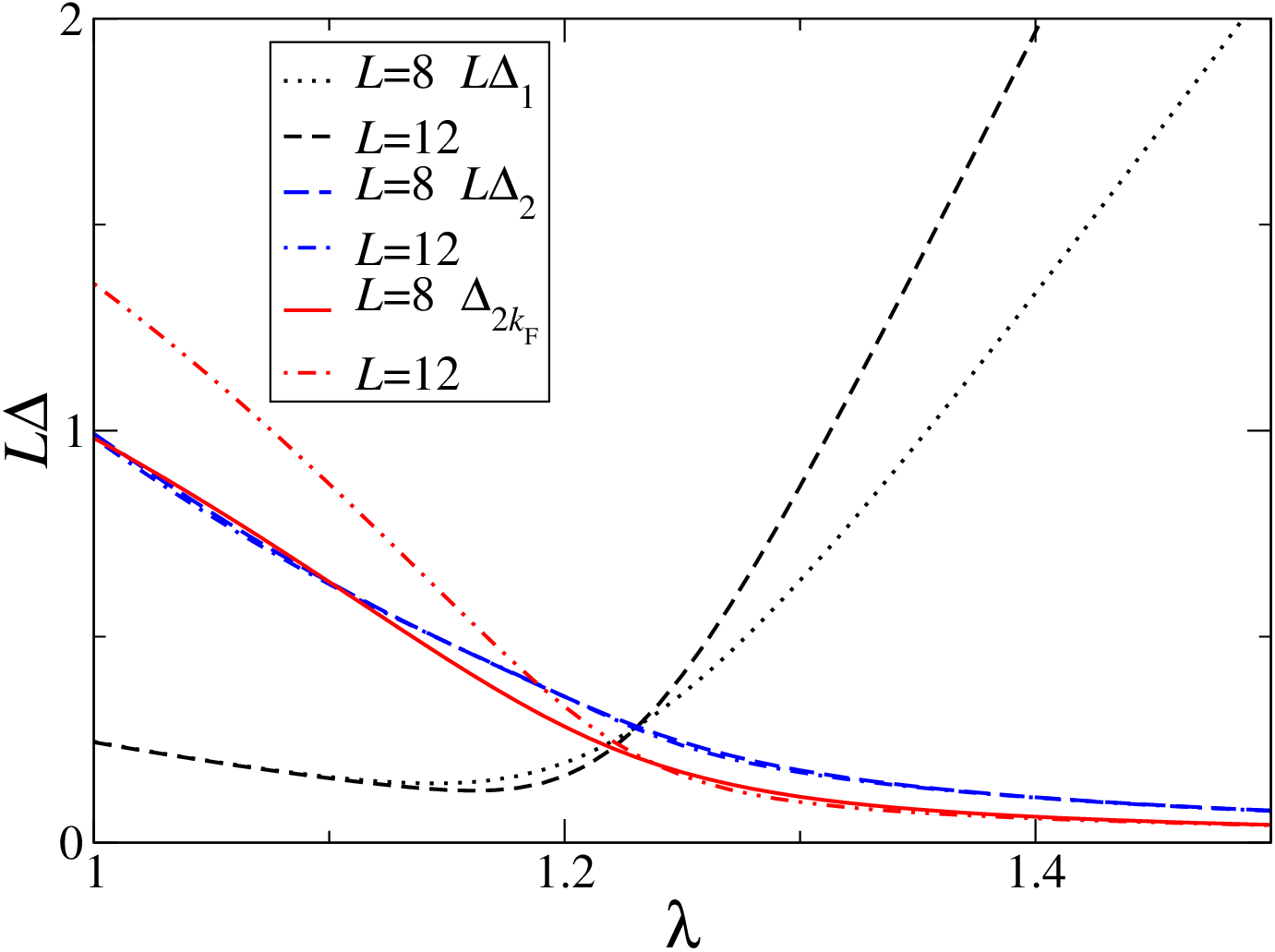}
         \caption{Scaled gaps $L\Delta_1$, $L\Delta_2$ and $L\Delta_{2k_{\rm F}}$ are plotted 
                  versus $\lambda$ for $L=8$ and $12$  for $m=1/2$. 
                           \par
                           ~\par
                           ~\par
                           ~\par}
         \label{sgmh}
      \end{center}
   \end{minipage}
   \hspace{0.04\columnwidth}
   \begin{minipage}{.48\linewidth}
      \begin{center}
        \includegraphics[scale=0.25]{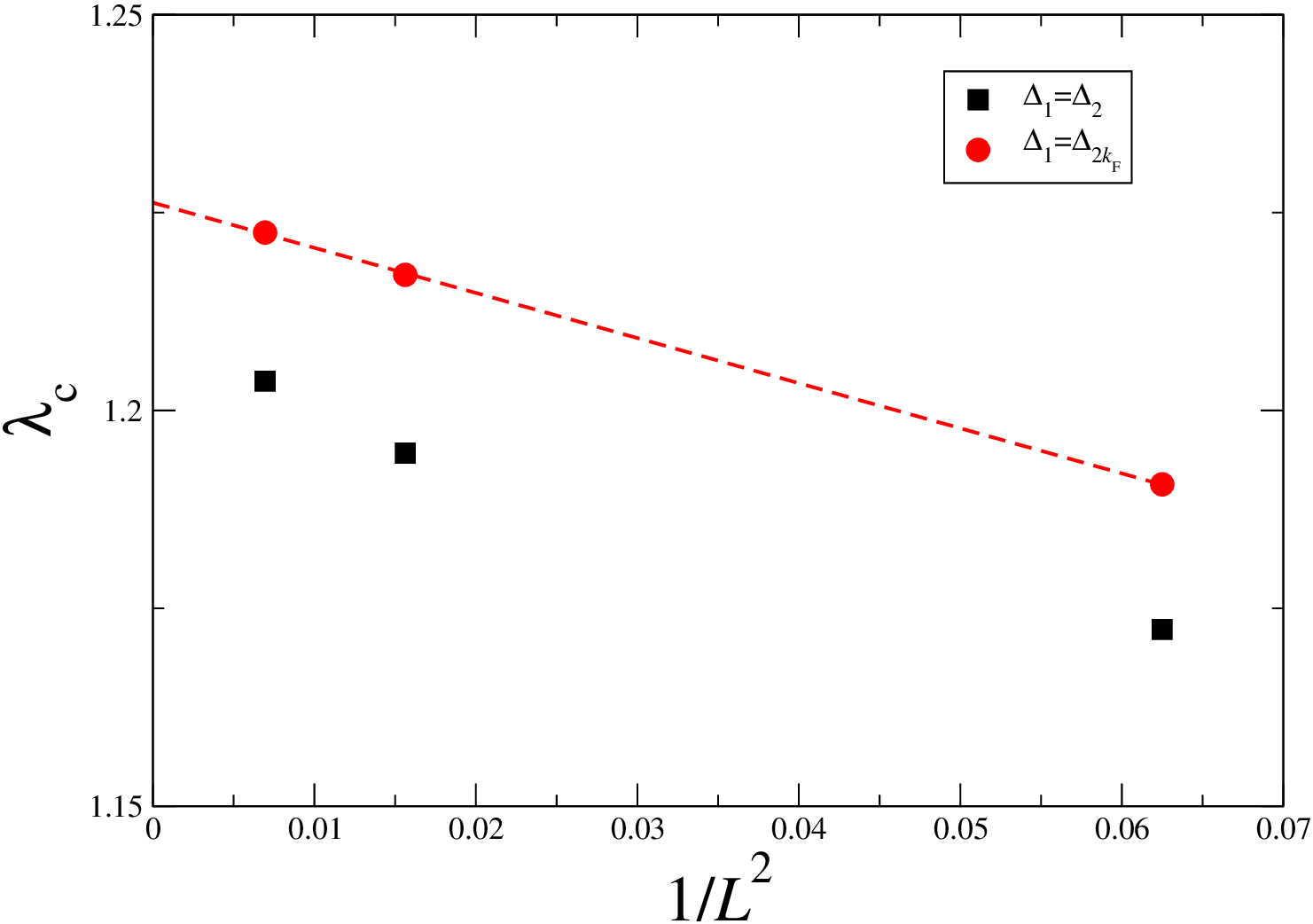}
        \caption{Extrapolation of the size-dependent fixed points $\lambda_{\rm c}$ 
                 of $L\Delta_1$ and $L\Delta_2$ (black squares), 
                 and those of $L\Delta_1$ and $L\Delta_{2k_{\rm F}}$ (red circles).
                 By use of the latter points, 
                 it results in $\lambda _{\rm c}= 1.2262 \pm 0.0003$ in the infinite length limit. 
           }
        \label{extrap2}
      \end{center}
   \end{minipage}
\end{figure}

In the magnetization process for $\lambda =1$ 
the system is in the CTLL phase for $0<m<1$. 
In the strong $J_1$ limit the spin pair on the $J_1$ bond 
forms the triplet with three states $|\uparrow  \uparrow \rangle$, 
$(|\uparrow \downarrow \rangle +|\downarrow \uparrow \rangle )/\sqrt{2}$ 
and $|\downarrow \downarrow \rangle$. 
For sufficiently large $\lambda$ two states  $|\uparrow  \uparrow \rangle$ 
and $|\downarrow \downarrow \rangle$ are stabilized and  
the remaining state $(|\uparrow \downarrow \rangle +|\downarrow \uparrow \rangle )/\sqrt{2}$ 
is excluded in the magnetization process. 
As a result the two-magnon bound state is realized and each 
magnetization step is not $\delta M$=1 but $\delta M$=2. 
This large $\lambda$ phase is also a kind of TLL phase, 
but different from the CTLL phase.
We call it the two-magnon TLL (TMTLL) phase. 
Now we consider three excitation gaps; 
they are the single-magnon excitation gap $\Delta _1$, the two-magnon excitation gap $\Delta _2$, and 
the $2k_{\rm F}$ excitation gap in the TMTLL phase $\Delta_{2k_{\rm F}}$. 
In both TLL phases $\Delta _2$ is gapless. 
In contrast, $\Delta _1$ ($\Delta _{2k_{\rm F}}$) is gapless (gapped) in the CTLL phase, 
but is gapped (gapless) in the TMTLL one. 
For $m=1/2$ the scaled gaps $L\Delta_1$, $L\Delta_2$ and $L\Delta_{2k_{\rm F}}$ are plotted 
versus $\lambda$ for $L=8$ and $12$  in Fig. \ref{sgmh}.  
The gapless and gapped behaviors of these excitations mentioned above are confirmed in Fig. \ref{sgmh}. 
The finite-size effect of the fixed points are shown in Fig. \ref{extrap2}.
The fixed points of $\Delta_1$ and $\Delta_{2k_{\rm F}}$ behave as $1/L^2$,
whereas those of $\Delta_1$ and $\Delta_2$ as $1/L$.
Unfortunately it is impossible to perform such extrapolations for general $m$.
Thus we use the cross point of $\Delta_1$ and $\Delta_{2k_{\rm F}}$
for largest $L$ as the phase boundary between the two TLL phases at each $m$,
which leads the difficulty in obtaining accurate boundary.
The error of the estimated boundary can be surmised  from Fig. \ref{extrap2}.

\section{Spin-Density-Wave and Nematic Spin Correlations}

\begin{figure}[h]
   \begin{minipage}{.48\linewidth}
      \begin{center}
         \includegraphics[scale=0.25]{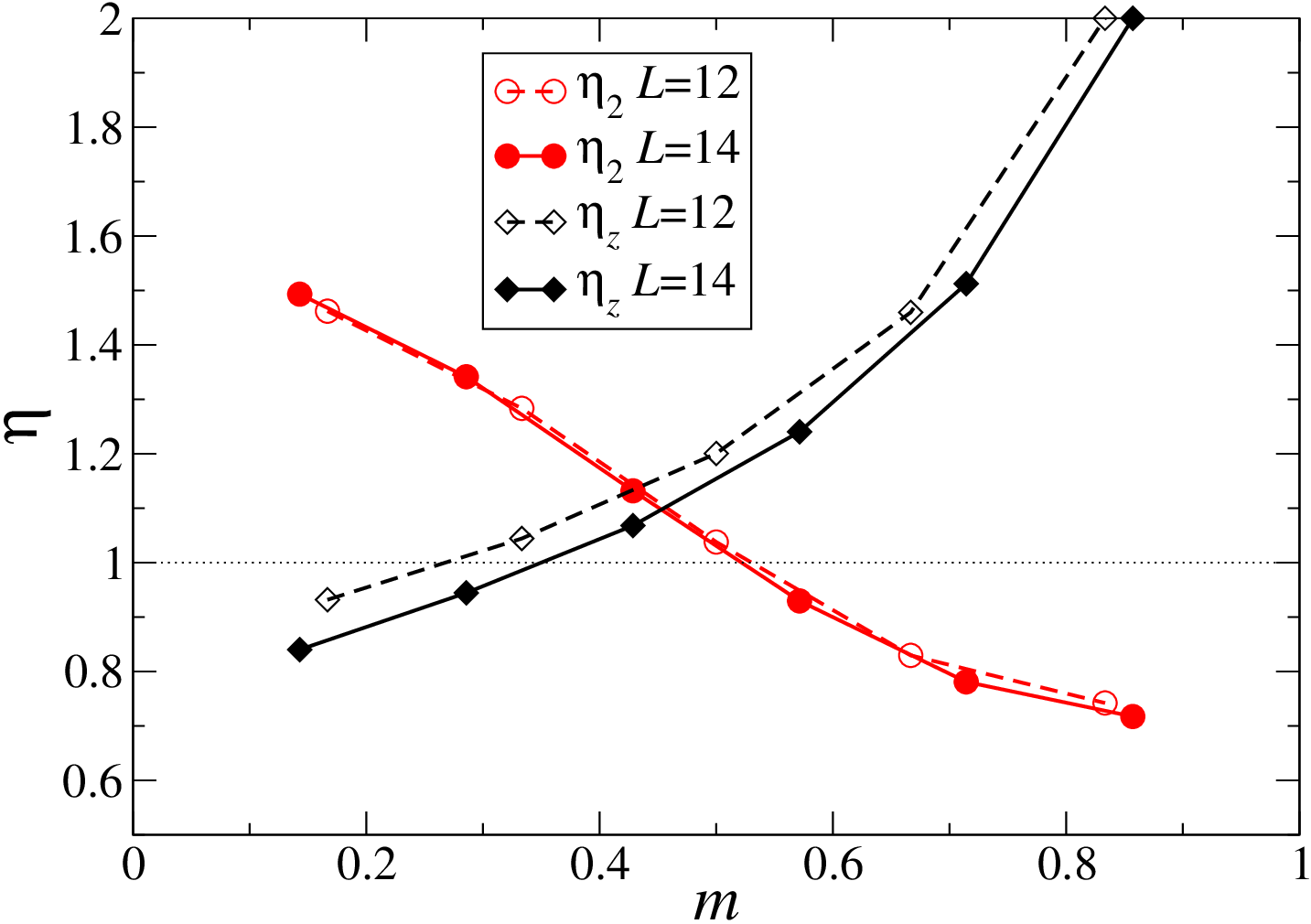}
         \caption{Exponents $\eta_2$ and $\eta_z$ estimated for $L=12$ and 14 are plotted versus 
                  $m$ for $\lambda =1.3$.
                           \par
                           ~\par
                           ~\par
                           ~\par}
         \label{eta}
      \end{center}
   \end{minipage}
   \hspace{0.04\columnwidth}
   \begin{minipage}{.48\linewidth}
      \begin{center}
        \includegraphics[scale=0.25]{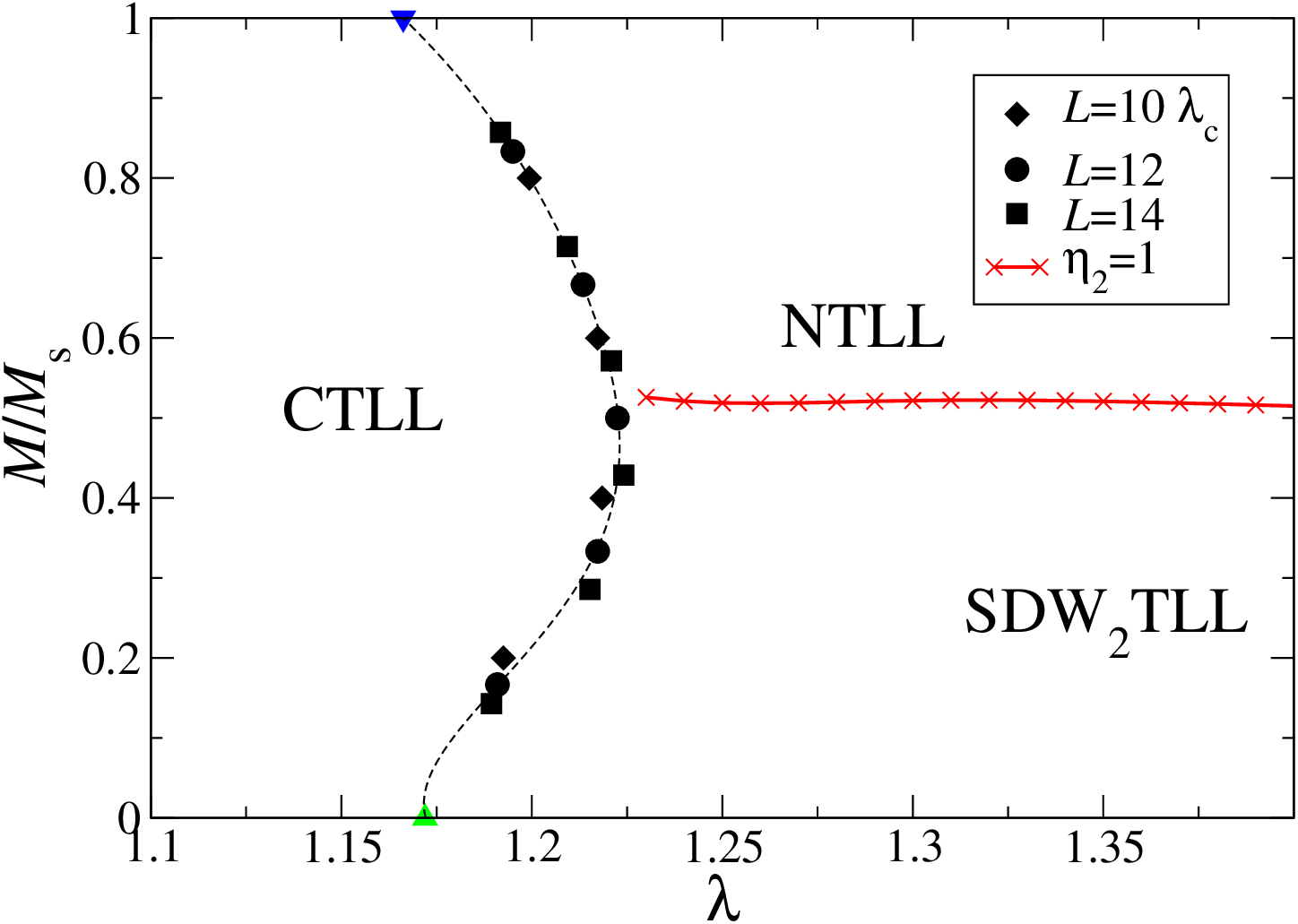}
        \caption{Phase diagram on the $\lambda-m$ plane for $J_2=0.3$. 
                 CTLL, SDW$_2$TLL and NTLL correspond to the conventional TLL, the SDW correlation dominant 
                 TLL and the spin nematic correlation dominant TLL phases, respectively. 
           }
        \label{mphase}
      \end{center}
   \end{minipage}
\end{figure}

The quasi-long-range spin-density-wave (SDW) and nematic orders are expected to appear in the 
TMTLL phases. 
They are characterized by the power-law decays of the following spin correlation 
functions
\begin{eqnarray}
&&\langle S^z_1 S^z_{2r+1} \rangle -\langle S^z \rangle ^2 \quad \sim \quad \cos (2k_{\rm F} r) r^{-\eta_z} ,
\label{sdw} \\
&&\langle S^+_1 S^+_2 S^-_{2r+1}S^-_{2r+2} \rangle \quad \sim \quad r^{-\eta _2}. 
\label{nematic}
\end{eqnarray}
Here Eq.(\ref{sdw})  corresponds to the SDW spin correlation parallel to the external field and 
Eq.(\ref{nematic}) corresponds to the nematic spin correlation perpendicular to the external 
field. 
The smaller exponent between $\eta_z$ and $\eta_2$ determines the dominant spin correlation. 
According to the conformal field theory these exponents can be estimated by the forms 
\begin{eqnarray}
\eta_2&=&{{E(L,M+2)+E(L,M-2)-2E(L,M)}\over{E_{k_1}(L,M)-E(L,M)}}, \\
\eta_z&=&2{{E_{2k_F}(L,M)-E(L,M)}\over{E_{k_1}(L,M)-E(L,M)}},
\label{exponent}
\end{eqnarray}
for each magnetization $M$, where $k_1$ is defined as $k_1=L/2\pi$. 
The exponents $\eta_2$ and $\eta_z$ estimated for $L=12$ and $14$ are plotted versus 
$m$ for $\lambda = 1.3$ in Fig.\ref{eta}.  
Since the system size dependence of $\eta _2$ is smaller than that of $\eta_z$,
we estimate the crossover point between the SDW dominant and the spin nematic dominant TLL phases 
as the point $\eta_2 =1$, assuming the relation $\eta_z \eta_2=1$
which should be satisfied in the TMTLL TLL phase.

\section{Phase Diagram and Magnetization Curve}

The phase diagram with respect to the anisotropy $\lambda$ and the magnetization 
for $J_2=0.3$ is shown in Fig. \ref{mphase}. 
The boundary between the CTLL and the TMTLL phases is 
estimated by $\Delta_1=\Delta _{2k_{\rm F}}$ for $L=10, 12$ and $14$. 
The crossover line between the spin-nematic dominant TLL (NTLL) and the SDW dominant TLL 
(SDW$_2$TLL) phases is estimated by $\eta_2=1$. 
We note that the TMTLL phase is composed of the NTLL phase and the SDW$_2$TLL phase.
The phase diagram indicates that the magnetization process for $\lambda \sim 1.2$ 
would meet two field-induced quantum phase transitions; one is between the SDW$_2$TLL and 
CTLL phases, the other is between the CTLL and NTLL ones. 
The magnetization curve calculated by the density matrix renormalization group (DMRG) for $L=48$ and $\lambda =1.2$ is 
shown in Fig. \ref{mag}. 
Since each magnetization step is $\delta M$=1 in the CTLL phase and $\delta M$=2 in the 
SDW$_2$TLL and NTLL regions, the two transitions are confirmed to occur.

\section{Discussion and Summary}

\begin{wrapfigure}[20]{r}{6cm}
    \vskip-1.0cm
    \centerline{\includegraphics[scale=0.30]{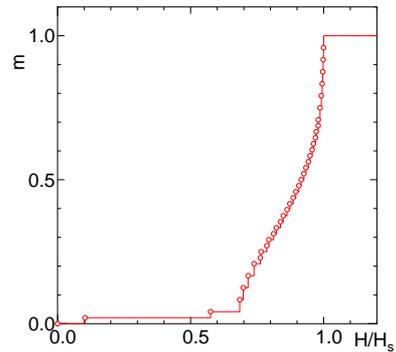}}
    \caption{
        Magnetization curve calculated by the DMRG for $L=48$, $J_2=0.3$ and $\lambda =1.2$. 
        Here $H_{\rm s}$ is the saturation magnetic field.
        We can see that the magnetization step is $\delta M =2$ in the low magnetization region, 
        $\delta M = 1$ in the intermediate magnetization region,
        and again $\delta M =2$ in the high magnetization region.}
        \label{mag}
\end{wrapfigure}
The mechanism for the appearance of the TMTLL phase in this model has been explained in \S2.
We think that the mechanisms are essentially the same for our previous the models
\cite{sakai2010,sakai2020,sakai2021,sakai2022,sakai2022a}. 
For instance, in the $S=1$ models \cite{sakai2020,sakai2021,sakai2022},
two states $|S^z=\pm 1\rangle$ of an  $S=1$ spin are selected
by the anisotropy effect, which is directly seen from the Hamiltonian.
This is very similar to the mechanism of the present model,
because $1-\lambda$ of the present model corresponds to the on-site anisotropy (so-called $D$ parameter)
of the $S=1$ model.
The TMTLL phase also appeared under the magnetic field
in the $S=1/2$ chain model with the nearest-neighbor ferromagnetic and the
next-nearest-neighbor antiferromagnetic interactions (NNF-NNNAF model) \cite{sudan,hikihara}.
The situation seems to be quite different in the NNF-NNNAF model,
because such a selection of states cannot be seen from the properties of the Hamiltonian itself.
We think that the appearance of the TMTLL phase in the  NNF-NNNAF model is
due to the combined many-body effect of the frustration and the magnetic field.
This reminds us that the dimerized state is realized in both of the $S=1/2$ bond-alternating antiferromagnetic
chain \cite{oka-sugi} and the $S=1/2$ antiferromagnetic chain with the next-nearest-neighbor interactions \cite{mg,oka-nomu}.
Their mechanisms are quite different from each other.
The former is explained by the bond-alternating nature of the Hamiltonian itself,
whereas the latter by the many-body effect originated from the frustration.

From the phase diagram Fig.\ref{mphase}, for $\lambda=1.2$,
we can estimate the lower critical field for the SDW$_2$TLL-CTLL boundary as
$m_{\rm c1} = 0.214$ and the upper one for the CTLL-NTLL boundary as $m_{\rm c2}=0.795$.
While, from Fig. \ref{mag},
we obtain $m_{\rm c1} = 0.208$ and $m_{\rm c2} = 0.708$.
The considerable difference in $m_{\rm c2}$ may come from the method
of the estimation of the phase boundary 
by spin gaps (as we stated, the size extrapolation is impossible in principle),
and also the steep curves of the phase boundary and the magnetization near $m_{\rm c2}$.
From the discussion on the behavior of correlation functions,
we said that the CTLL-TMTLL boundary can be determined by
$\Delta_1 = \Delta_2$ or $\Delta_1 = \Delta_{2k_{\rm F}}$
(actually we used the latter for drawing the phase diagram).
On the other hand,
the condition for the change of the magnetization step between $\delta M=1$ and $\delta M=2$
is $2\Delta_1 = \Delta_2$, which is different from the above condition $\Delta_1 = \Delta_2$.
This seeming contradiction can be resolved as follows.
Since $\eta_2 = 4\eta_1$ in the CTLL phase \cite{hikihara},
the ratio of the spin gaps $\Delta_2/\Delta_1$ is 4 in the limit of $L \to \infty$.
In the TMTLL phase,
$\Delta_1$ is gapped and $\Delta_2$ behaves as $1/L$,
which leads to $\Delta_2/\Delta_1 =0$ in the limit of $L \to \infty$.
Thus, $\Delta_2/\Delta_1$ jumps from 4 to 0 at the CTLL to TMTLL transition point in this limit.
Therefore both of the condition $\Delta_1 = \Delta_2$ and $2\Delta_1 = \Delta_2$
converge into the same CTLL to TMTLL transition point in  the thermodynamical limit.

In summary, 
the $S=1/2$ ferromagnetic and antiferromagnetic bond-alternating chain with the coupling anisotropy 
at the ferromagnetic bond is investigated using the numerical diagonalization. 
For sufficiently large Ising-like anisotropy the field-induced NTLL
phase appears as well as the SDW$_2$TLL one. 
The phase diagram with respect to the anisotropy and the magnetization for a typical 
parameter is presented. 
The behavior of the magnetization curve by the DMRG is consistent with the phase diagram.

\section*{Acknowledgment}
This work has been partly supported by JSPS KAKENHI, Grant Numbers 16K05419, 
16H01080 (J-Physics), 18H04330 (J-Physics), JP20K03866, and JP20H05274. 
We also thank the Supercomputer Center, Institute for Solid State Physics,
University of Tokyo and the Computer Room, Yukawa Institute for Theoretical
Physics, Kyoto University for computational facilities.
We have also used the computational resources
of the supercomputer Fugaku provided by the RIKEN
through the HPCI System Research projects (Project ID:
hp200173, hp210068, hp210127, hp210201, and hp220043).

\end{document}